\documentstyle[12pt,epsfig]{article}

\def\pr{{\it Phys. Rev.}}
\def\prl{{\it Phys. Rev. Lett.}}
\def\pl{{\it Phys. Lett.}}

\def\grg{{\it Gen. Relativ. Grav.}}

\def\apj{{\it Ap. J.}}

\begin{document}
\begin{titlepage}

\title{Cosmological consequences of an inhomogeneous space-time}

\author{{Fabrizio Canfora\thanks{canfora@sa.infn.it},
 Antonio Troisi\thanks{antro@sa.infn.it}}
\\ {\it Dipartimento di Fisica``E. R. Caianiello''} \\
\it Universit\`{a} di Salerno, Via S. Allende-84081 Baronissi (SA), Italy \\
 {\it Istituto Nazionale  di Fisica Nucleare, sez. di Napoli }\\{\it Gruppo
Collegato di Salerno}}

\date{\today}

\maketitle

\begin{abstract}
Astrophysical observations provide a picture of the universe as a
4-dim homogeneous and isotropic flat space-time dominated by an
unknown form of dark energy. To achieve such a cosmology one has
to consider in the early universe an inflationary era able to
overcome problems of standard cosmological models.\\
Here an inhomogeneous model is proposed which allows to obtain a
Friedmann-Robertson-Walker behaviour far away from the
inhomogeneities and it naturally describes structures
formation. \\
We also obtain that the cosmological term does not prevent
structure
formation, avoiding a fine tuning problem in initial conditions.\\
The asymptotic exact solution have been calculated. A simple test
with universe age prediction has been performed. A relation
between the inhomogeneity, the breaking of time reversal, parity
and the matter-antimatter asymmetry is briefly discussed.

\end{abstract}

\vspace{1.7cm}

\small{{\bf keywords}:inhomogeneous cosmology, structures
formation, matter antimatter asymmetry}
\vspace{0.3cm}
\\
\small{PACS
number(s) {98.80.Cq, 98.80.Bp, 04.20.Jb}}

\end{titlepage}

\section{Introduction}

Today, thanks to new refined experimental techniques (i.e.
neoclassical tests, see \cite{peebles}) the experimentalists can
explore a very large redshift region of the universe. In
particular, Supernovae Cosmology Project (SCP) \cite{perlmutter},
High-Z search Team (HZT) \cite{riess} have been able to analyze
supernovae data at very high redshift providing an astonishing
test able of discriminate between different cosmological models.
On the other hand, cosmic microwave surveys
\cite{cobe,boomerang,maxima,wmap} have investigated the universe
background until the {\it last scattering surface}, giving
significant indications on the geometry of the universe. The
overall result is the indication of a cold dark matter universe
dominated by an unknown dark energy ($\Lambda$-CDM model), which
is featured as a spatially flat Friedmann-Robertson-Walker (FRW)
manifold from
the geometrical point of view.\\
These great improvements in experimental cosmology represent a
challenge for the various theoretical models, which result
severely constrained by observational data.\\ It is a remarkable
fact that the best fit model for the dynamical evolution of the
universe according to the new experimental data, is a flat FRW
cosmology. From the theoretical point of view, the fact that the
universe is homogeneous and isotropic to a very high degree of
precision rises several problems. The inflationary scenario,
thanks to a very fast expansion of early universe which smooths
out the inhomogeneities, is the standard way to explain the
genesis of such a FRW cosmology (although it is not the unique
way, see, for example, \cite{holland}).\\
However, there is still not a commonly accepted model among the
various inflationary proposals. For these reasons, it could be
useful to study the non-linear evolution of the inhomogeneities
unlike the standard case which uses the linearized theory. Here,
we propose a relativistic cosmological model which could explain
how the inhomogeneities die off during the expansion of the
universe and can be developed the
structures formation.\\
Our framework is an inhomogeneous 4-dim space-time of
phenomenological origin. The study of an inhomogeneous
cosmological model has a very long history \cite{krasinsky}.
However, in the existing literature on this subject it is not easy
to make contact with the observations and, in particular, to
answer to the question about how the inhomogeneities evolve in
time and why the observed universe is so near to a flat FRW. A
main attempt to study the cosmological consequences of an
inhomogeneous space-time has been developed in 1967 by P.J.E.
Peebles \cite{peebles0}, deepening the cosmological model based on
Tolman-Bondi metric \cite{t-b}. The author, in particular,
obtained interesting results on structures formation starting from
a spherically symmetric inhomogeneity in absence of cosmological
constant.
\\
Another interesting approach is the so-called ``swiss-cheese"
cosmological model \cite{swiss-cheese}. This scheme, originated by
Lemaitre suggestions was developed by Einstein to study the
effects of universe expansion on the solar system \cite{ellis}.
Later on, the properties of this model have been exploited from a
cosmological and astrophysical point of view. In such an approach
spherical inhomogeneities described by local Schwarzschild or
Tolman-Bondi metrics are embedded in a FRW manifold by the means
of suitable matching conditions. Several improvements have been
performed in time to this framework \cite{ellis,swiss-cheese2}.
Some considerations on its observational
effects have been furnished in \cite{swiss-cheese2}.\\
As we will see the model proposed here bears some resemblance with
the {\it swiss-cheese} scheme. However, we will not impose any
matching condition and the metric near the inhomogeneity is
computed by only looking to at the field equations near the
inhomogeneity itself. In this way we will reduce the freedom of
the standard swiss-cheese approach. Nevertheless, the quite
general result is that, near the inhomogeneity, the metric behaves
like a self-similar fluid, thus providing a physical basis to the
swiss-cheese models themselves.\\
The results, derived from the exact non vacuum Einstein equations
with a dust source, show that, far away the inhomogeneity the
evolution of the universe is almost a standard flat FRW while the
size of the inhomogeneity does decrease in time. This behaviour is
quite according with the results obtained in \cite{peebles0}. In
this context the problem of the structures formation could be
accounted in a natural way. The model also reveals an intriguing
relation between the smoothing out of the inhomogeneity and the
breaking of the time reversal and of the parity.\\ A further
analysis is performed with cosmological constant. The result is a
coherent dynamical universe in which structure formation develops
in a natural way. Cosmological constant dominates for greater time
and induces a de Sitter
expansion far away the inhomogeneities.\\
The work is organized as follows: in Sect.II we propose the model.
Sect.III is devoted to study of of the exact asymptotic behaviour.
In Sect.IV we provide the analysis in presence of cosmological
constant. Sect.V contains a simple test with universe age
prediction. Sect.VI is devoted to a summary and to some
consideration on the topic.

\section{The model}

The first problem to construct the model is how to describe an
inhomogeneity. On a FRW background, since the metric is
homogeneous, the best thing one can do is to study the evolution
of the perturbations in the energy density with the linearized
Einstein equations. There is an extensive literature on this
subject (see for example, \cite{nonhomocosmologies} and the
references therein) and we will not enter in these details.
However, the nonlinearities are the most characteristic features
of the Einstein equations. Hence, some physical implications of
the theory, that manifest themselves in the full Einstein
equations, could be lost in the linearized
ones. Thus it is interesting to try to use the exact equations.\\
Following the observations, we will try to describe a flat FRW
model with an inhomogeneity. As source we will take simply a dust
energy-momentum tensor:
\begin{equation}
T_{\mu \nu }=\rho u_{\mu }u_{\nu }\,,
\end{equation}
where the energy density $\rho $ does depend on the radial coordinate too: $%
\rho =\rho (t,r)$.\\
We will take the metric in the form:
\begin{eqnarray}
g_{\mu \nu }dx^{\mu }dx^{\nu }
&=&ds^{2}=dt^{2}-\frac{2K}{A}dtdr-A^{2}\left[
dr^{2}+r^{2}d\Omega \right] ,  \label{metrica} \\
A &=&A(t,r),\ \,K=const>0\,\,. \nonumber
\end{eqnarray}
We make this ans\"{a}tz to careful describe the inhomogeneity. The
$r-$dependence of the scale factor $A$\ takes into account the
radial deformations of the $\Sigma _{t}$ ($t=const$) hypersurfaces
which slice space-time in FRW metric. The nondiagonal term in the
metric $g_{tr}=-2K/A$ gives a radial component to the dust
velocity field:
\begin{equation}\label{3}
g_{tr}\neq 0\Rightarrow u_{r}\neq 0.
\end{equation}
The particular form chosen for $g_{tr}$ will allow to find, in the
limit $A\gg 1$, solutions approaching to a FRW metric and, in such
a case, $t$ becomes the physical co-moving time. Of course, it is
always possible to diagonalyze the metric (\ref{metrica}) with a
suitable coordinates transformation and in such a case one obtains
a slight generalization of the ans\"{a}tz considered in
\cite{peebles0}. However, in the present case, the diagonalized
form of the new timelike coordinate\footnote{Roughly speaking when
we diagonalyze metric (\ref{metrica}) it appears a non trivial
coefficient in front of the $(dt^\prime)^{2}$.} cannot be related
anymore in a direct way to the cosmic time so that, in the new
coordinates
system, the physical interpretation is less transparent.\\
We could choose a different form for $g_{tr}$, but, among
the various natural choices (such as $\displaystyle{g_{tr}\sim \frac{1}{A^{n}}}$ with $n>1$%
, or $g_{tr}\sim \exp (-A)$), this simplifies somehow the
calculations. Finally, the constant $K$, that, for sake of
simplicity could be put equal to one, is useful to recognize the
role of $g_{tr}$.

\noindent Now, the explicit expressions of the components of the
Einstein equations are not particulary expressive. Thus, we write
down only the main equation we have to solve:

\begin{displaymath}
G_{33}=0\Rightarrow
\end{displaymath}
\vspace{0.05cm}
\begin{eqnarray}
rA^{3}\left( 5K^{2}+A^{4}\right) \left( \partial _{t}A\right)
^{2}+rK\left( A^{4}-K^{2}\right)
\partial _{t}A\partial _{r}A-K^{3}2A\partial _{t}A+ \nonumber\\ \label{equazione33}
+rA\left( A^{4}-K^{2}\right) \left( \partial _{r}A\right)
^{2}-A^{2}\left( 3K^{2}+A^{4}\right) \partial _{r}A+2rA^{4}\left(
K^{2}+A^{4}\right) \partial _{t}^{2}A+
\\ \nonumber-KrA\left( K^{2}+A^{4}\right) \partial
_{tr}^{2}A-rA^{2}\left( K^{2}+A^{4}\right) \partial _{r}^{2}A
=0\,,
\end{eqnarray}
\vspace{0.3cm}

\noindent the other equations (related to the other
components of Einstein tensor) determine simply the explicit form
of energy-momentum tensor once the scale-factor $A$ has been
calculated by
(\ref{equazione33}).\\
Eq.(\ref{equazione33}) is highly nonlinear and it cannot be solved
by separation of variables. However, its asymptotic behaviour
allows to clarify some important
physical aspects of the problem.\\
In fact, this equation tells us that it is natural to separate the
evolution for big $r$ (i.e. far away the inhomogeneity) from the
evolution for small $r$ (i.e. on the ''top'' of the
inhomogeneity). It is clear that for $r\rightarrow \infty $ the
terms multiplied by $r$ will dominate, while for $r\rightarrow 0$
the terms without $r$ will dominate.\\
As an important remark we stress that Eq.(\ref{equazione33}) is
not scale invariant. Namely, if $A$ is a solution then $\lambda A$
(with constant $\lambda $) cannot be a solution. This means that
due to the $g_{tr}$ coefficient
there is a characteristic scale in the dynamics of $A$ which we will call $%
A_{Crit}\sim K^{\frac{1}{2}}$.\\
Another interesting consequence of the model, related to the
nondiagonal term in the metric, is that the dynamics is not
invariant under the time reversal $T$: $t\rightarrow -t$ and under
the parity $P$: $r\rightarrow -r$ (although it is invariant under
$TP$). This fact could be very interesting if $t$ would be the
physical comoving time, but $t$ is not. However, if one finds a
FRW-like solution, in this limit, $t$ is the physical comoving
time and one is allowed to relate, as we will see, this breaking
of $T$ and $P$ to the
evolution of the inhomogeneity.\\
Previous considerations indicates some differences between our
model and swiss-cheese ones. In swiss-cheese models one imposes by
hand the matching conditions needed to join the internal
inhomogeneous space-time (such a Schwarzschild or a self-similar
fluid) to the external FRW space-time. This approach results in a
non-smooth metric at the embedding hyper-surface. On the other
side in our scheme it is possible to study the general behaviour
of the space-time near and far-away the inhomogeneity without any
matching condition. As a consequence this approach allows to
define observable quantities, i.e. the age of universe, in term of
parameters characterizing the
intrinsic size of the inhomogeneity.\\
It is worth to note that it is impossible to shed light on the
relation between the breaking of T and P and the inhomogeneity in
the standard swiss-cheese models.

\section{Asymptotic behaviour}

As remarked in Sec.1, because of the complexity of
Eq.(\ref{equazione33}), the model solution will be studied only in
the asymptotic cases. We start by analyzing the behaviour far away
from the inhomogeneity, as second case we propose the near limit.

\subsection{Far away the inhomogeneity: a FRW-like evolution}

The model behaviour far away the inhomogeneity can be achieved in
the limit $r\rightarrow \infty$. In this case,
Eq.(\ref{equazione33}) becomes:
\begin{eqnarray}\label{far}
A^{3}\left( 5K^{2}+A^{4}\right) \left( \partial _{t}A\right)
^{2}+K\left( A^{4}-K^{2}\right) \partial _{t}A\partial
_{r}A+\nonumber\\+A\left( A^{4}-K^{2}\right) \left( \partial
_{r}A\right) ^{2} +2A^{4}\left( K^{2}+A^{4}\right) \partial
_{t}^{2}A+\\-KA\left( K^{2}+A^{4}\right) \partial
_{tr}^{2}A-A^{2}\left( K^{2}+A^{4}\right) \partial _{r}^{2}A=0
\nonumber \,.
\end{eqnarray}

The model, to have the physical interpretation of an universe with
an inhomogeneity only at small $r$, must admit solutions that do
not depend on $r$ in the limit
$r\rightarrow \infty $. \\
Eq.(\ref{far}) has a solution depending only on $t$. Indeed,
writing $A(t,r)=A(t)$ it becomes:
\begin{equation}
\left( 5K^{2}+A^{4}\right) \left( \partial _{t}A\right)
^{2}=-2A\left( K^{2}+A^{4}\right) \partial _{t}^{2}A\,.
\end{equation}
As a consequence $A$ is implicitly given by the following
expression:

\begin{eqnarray}
I_{1}t+I_{2} &=&\int^{A^{\prime
}}\sqrt{\frac{x^{5}}{1+x^{4}}}dx\,,
\\\nonumber A^{\prime } &=&\frac{A}{K^{\frac{1}{2}}} \ \ \
I_{i}=const\,,
\end{eqnarray} where $I_{i}$ are integration constants.\\
It is possible to verify that for great values of time such a
solution reduces to $A(t)\sim{t^{2/3}}$. This result is
intriguing, according with a FRW evolution
for dust matter as source.\\
As a consequence, we are allowed to interpret $t$ as the
asymptotic comoving time, without
loss of generality. This fact will be relevant in Sect.V.\\
Moreover such a result is quite natural since we have not imposed
any matching condition unlike swiss-cheese models.

\subsection{On the top of the inhomogeneity: a decreasing scale
factor}

Now, we will study the dynamics near the inhomogeneity, i.e. in
the limit $r\rightarrow 0$. In this case Eq.(\ref{equazione33})
becomes:
\begin{equation}\label{near}
-K^{3}2\partial _{t}A-A\left( 3K^{2}+A^{4}\right) \partial
_{r}A=0\,.
\end{equation}

\noindent It is important to stress here that, with $g_{tr}=0$
(i.e. $K=0$), the above equation would become trivial. Then, the
importance of the role of the
inhomogeneity manifests itself for small $r$. Eq.(\ref{near}) has no $r-$%
independent solution and, moreover, cannot be solved by separation
of variables. Nevertheless, it is possible find an interesting
exact solution.
As an intermediate step, let us study firstly the two extreme cases: \textit{%
i)} $A\gg K^{\frac{1}{2}}$, \textit{ii)}$A\ll K^{\frac{1}{2}}$.

\subparagraph{\textit{i) }$A\gg K^{\frac{1}{2}};$} In this case
the Eq.(\ref{near}) reads:
\begin{equation}
-K^{3}2\partial _{t}A-A^{5}\partial _{r}A=0\,,
\end{equation}
and it can be now solved by separation of variables. Taking $A(t,r)=%
\overline{a}(t)\overline{b}(r)$, we obtain the following solution:
\begin{eqnarray}
\overline{a}(t) &=&K^{\frac{1}{2}}\left( d_{1}+\frac{5}{2}\sigma \frac{t}{K^{%
\frac{1}{2}}}\right) ^{-\frac{1}{5}} \nonumber  \\
\\\overline{b}(r) &=&\left( d_{2}+5\sigma r\right) ^{\frac{1}{5}}\,, \nonumber
\end{eqnarray}
where the $d_{i}$ are integration constants and $\sigma $ is the
separation constant. We get the interesting result that, on the
top of the inhomogeneity the scale factor is a decreasing function
of $t$, the part of the solution depending by $r$ approaches to
constant for $r\rightarrow{0}$.

\subparagraph{\textit{ii) }$A\ll K^{\frac{1}{2}};$}

In this other case Eq.(\ref{near}) becomes:
\begin{equation}
-K2\partial _{t}A-3A\partial _{r}A=0\,.
\end{equation}
Also in this case it is possible to separate the variables. By
taking again $A(t,r)=\overline{a}(t)\overline{b}(r)$, one shows
that:
\begin{eqnarray}
\overline{a}(t) &=&K^{\frac{1}{2}}\left( d_{3}+\frac{1}{2}\sigma ^{\prime }%
\frac{t}{K^{\frac{1}{2}}}\right) ^{-1} \nonumber \\ \\
\overline{b}(r) &=&d_{4}+\frac{1}{3}\sigma ^{\prime }r \,,
\nonumber
\end{eqnarray}
where again the $d_{i}$ are integration constants and $\sigma
^{\prime }$ is the separation constant. In this case also one
obtains that on the top of the inhomogeneity the scale factor, and
then the size of the inhomogeneity, is a decreasing function of
$t$, while the radial part of the solution tends to a constant
value.

It is useful to note that in both cases, if $A$ depends on
$r^{\alpha }$ (with $\alpha >0$) then $A$ depends on $t^{-\alpha
}$. Hence, it is natural to try to find a solution of the full
equation (\ref{near}) in the form $A=A(r/t)$. By substituting this
ans\"{a}tz in the (\ref{near}), one immediately gets the following
implicit expression of $A$ as function of $r/t$:
\begin{equation}
\frac{A}{K^{\frac{1}{2}}}\left( 3+\left(
\frac{A}{K^{\frac{1}{2}}}\right) ^{4}\right)
=2\frac{K^{\frac{1}{2}}r}{t}\,.
\end{equation}
A few remarks on this solution are in order.\\
First, $A$ is not factorized and has a true singularity for $r=0$
(there the Ricci scalar diverges). On the other hand, it
interpolates between the two extreme behaviors: for large $t$ one
has $A\sim \left(r/t\right)$ while for small $t$ one has $A\sim
\left(r/t\right) ^{\frac{1}{5}}$. Moreover, $A$ is always a
decreasing function of $t$, this implies that the size of the
inhomogeneity does decrease with time and at the same time the
matter density increases.
 \\
It is important to stress that the term in the equation that
implies such an effect, i.e. $-K^{3}2A\partial _{t}A$, is strictly
related to the non-diagonal term: in fact, with $g_{tr}=0$ this
term would be
zero.\\
It has to be remarked that, if one uses the linearized Einstein
equations and treats the non-diagonal term as a small
perturbation, then this effect disappears, since it is of the
third order in $K$. The decreasing of the size of the
inhomogeneity could explain in a natural way the structures
formation (see also \cite{peebles0}). In fact, a decreasing scale
factor implies an increasing density $\rho$. Thus, one would
expect that, when the density reaches a critical value $\rho
_{Crit}$ \footnote{We stress that with $\rho_{Crit}$ we are
referring to the critical matter density needed to allow gravity
to form structures. Obviously this definition must be carefully
distinct from the analogous expression with which is characterized
the energy density amount able to provide a spatially flat
universe}. (that depends on the actual cosmological structure we
are considering, such as galaxies, cluster of galaxies, etc.),
then the structure begins the formation decoupling from the
outside universe that evolves as in the usual FRW models. Of
course, when $\rho
>\rho _{Crit}$, one has to consider the contribution of the
pressure to the energy-momentum tensor.\\
Since in this case the integral curves of $\partial _{t}$ are not
geodesic, this model could suggest an interpretation of the
observed acceleration of the galaxies without the introduction of
an extra scalar field. In fact, if the galaxies would follow the
integral line of $\partial_{t}$, then these would accelerate with
respect to each others. Thus the observed negative deceleration
parameter could be simply explained as an effect of geometrical
origin.
\\
Eventually, we want to comment the symmetry breaking. In this
model, the smoothing out of the inhomogeneity and the breaking of
$T$ and $P$ are strictly related. Both phenomena have the same
physical origin: the non-diagonal term. In other words, one cannot
obtain the first without the second and viceversa. In the standard
perturbative approaches to the study of the cosmological
anisotropies, one cannot see this relation because the FRW
background is $T$ and $P$ invariant. Only the full theory can
reveal this interesting connection. This could have important
phenomenological consequences for the matter-antimatter asymmetry.
In fact, although almost all the calculations of particles
creation by time-dependent gravitational field predict an equal
number of particles and anti-particles, it is well known that a
violation of the homogeneity or of the parity can break this
symmetry (see, for example, \cite{gibbons}) (even in the paper of
Parker \cite{parker}, one of the first work on the subject, it is
stressed that the fact that the particles and antiparticles are
created in pair is related to the homogeneity). Moreover, it has
been shown by Klinkhamer \cite{klinkhamer} that even gauge theory
on flat background, but with a nontrivial topology, can have CPT
breaking and in our case, since we have a
singularity for $r=0$, the topology of the spacelike slices is $%
R^{3}\setminus \left\{ 0\right\} $. So, this kind of effect could
have important consequences in the study of the matter-antimatter
asymmetry problem.

\section{The model with cosmological constant}

In this section we analyze the model in presence of cosmological
constant. \\
The field equations, as it is well known, read:

\begin{equation}
G_{\mu\nu}=T_{\mu\nu}+\Lambda g_{\mu\nu}.
\end{equation}

\noindent By these, inserting the metric described in relation
(\ref{metrica}), we obtain the 3-3 component which determines the
cosmological dynamics:

\begin{displaymath}
G_{33} =\Lambda g_{33}\Rightarrow
\end{displaymath}

\begin{eqnarray}
&&rA^{3}\left( 5K^{2}+A^{4}\right) \left(
\partial _{t}A\right) ^{2}+rK\left( A^{4}-K^{2}\right)
\partial _{t}A\partial _{r}A+ 2rA^{4}\left( K^{2}+A^{4}\right)
\partial _{t}^{2}A+ \nonumber \\&& \label{equazione33L}
+rA\left( A^{4}-K^{2}\right) \left( \partial _{r}A\right)
^{2}-A^{2}\left( 3K^{2}+A^{4}\right) \partial
_{r}A-K^{3}2A\partial _{t}A + \\&& \nonumber -KrA\left(
K^{2}+A^{4}\right) \partial _{tr}^{2}A-rA^{2}\left(
K^{2}+A^{4}\right) \partial_{r}^{2}A =-\Lambda
rA\left(A^{4}+K^{2}\right)^{2}.
\end{eqnarray}

Now, we can study, again, the model far away and near the
inhomogeneity.\\

In the first case we consider $r$ $\rightarrow\infty$ and a
$r$-independent solution, so that $\partial_{r}{A}\rightarrow{0}$.
The result is a simple generalization of the previous case:

\begin{equation}
2K^{2}A^{3}\partial^{2}_{t}{A}+5K^2{A}^{2}(\partial_{t}{A})^{2}
+A^{6}(\partial_{t}{A})^2+2A^{7}\partial^{2}_{t}{A}=-\Lambda(A^4+K^2)^2
\end{equation}

\noindent which can be rearranged in a more expressive form

\begin{equation}\label{far-L}
-\Lambda{(1+\frac{K^2}{A^4})}^{2}=2K^{2}\frac{\partial^{2}_{t}{A}}{A^{5}}+5K^2\frac{(\partial_{t}{A})^{2}}{{A}^{6}}
+\frac{(\partial_{t}{A})^2}{A^{2}}+2\frac{\partial^{2}_{t}{A}}{A}\,.
\end{equation}

\noindent It is clear that, for great values of $A$, this relation
becomes the pressure Friedmann equation for the standard cosmology
(i-i component of Einstein equation in FRW-metric) in presence of
cosmological constant \cite{peebles,kolb}; we recall that we have
assumed a pressureless matter fluid, so that we obtain:

\begin{equation}
-\Lambda=\frac{(\partial_{t}{A})^2}{A^2}+2\frac{\partial_{t}^2{A}}{A}\,.
\end{equation}

\noindent It can be shown that Eq.(\ref{far-L}) has a solution
which, for $t\rightarrow\infty$, agrees with the requested de
Sitter expansion for a FRW spatially flat model with cosmological
constant. In other words, we achieved an asymptotically de Sitter
behaviour
as a natural effect of the model for great values of $r$.\\
\\
Let us we study our model near the inhomogeneity (i.e. in the
limit ($r\rightarrow{0}$)). Eq.(\ref{equazione33L}) gives:

\begin{equation} \label{near-L}
\frac{1}{A^{4}}\left(3A^{2}K^{2}\partial_{t}{A}+2AK^{3}\partial_{t}{A}+A^{6}\partial_{r}A\right)=\Lambda
Ar\,.
\end{equation}

\noindent For $r\rightarrow 0$ the right term fall down and
(\ref{near-L}) becomes the same as in the case without
cosmological constant

\begin{equation}
\left(3A+2K\right)K^{2}\partial_{t}{A}+A^{5}\partial_{r}A=0\,.
\end{equation}

\noindent Thus, a cosmological constant term inside this
inhomogeneous cosmological model does not destroy the
inhomogeneity and, more important, this term does not influence
the evolution of such an inhomogeneity. Such a behaviour is not
trivial. Unlike in the standard theory (see
\cite{nonhomocosmologies} and references therein) in which the
$\Lambda$-term can prevent structures formation, in this model
structure formation will be preserved in presence of cosmological
constant. Furthermore, it is not required any sort of fine tuning
between matter and cosmological component to allow to cosmological
perturbations of determine today observed structures.

\section{A simple test}

\noindent To simply test our model, we have checked its capability
of
providing a significant prediction of the age of the universe.\\
To perform this test we use the exact solution obtained far away
the inhomogeneity by which is significant the use of $t$ as comoving time.\\
The scale factor was determined by the expression:

\begin{equation}\label{test1}
I_{1}t+I_{2}=\int^{A^\prime}\sqrt{\frac{x^5}{1+x^4}}dx\,,
\end{equation}

\noindent in this limit.\\
In this model, the age of the universe is directly related to the
ratio between the actual size of the universe and the
characteristic size of the inhomogeneity ($K^{1/2}$).
\\
It is always possible to assume $I_2=0$, thus Eq.(\ref{test1}) can
be written as:

\vspace{7mm}
\begin{equation}\label{test-time}
t=\frac{1}{I_{1}}\int^{A^\prime}\sqrt{\frac{x^5}{1+x^4}}dx\,.
\end{equation}
\vspace{7mm}

\noindent It is worth to note that the predicted age of the
universe results slightly smaller than the estimate for standard
FRW and swiss-cheese models(in particular, the greater is the size
of the inhomogeneity the smaller is the predicted age
\footnote{The smaller value of age prediction for our model
descends by the presence of the sum in the denominator of right
member of (\ref{test-time}). In the standard case this term would
be simply $x^{4}$.}). We can compute the value of $A^{\prime}$ at
today by taking the current time deduced by observations. We will
take
$I_{1}$ as the measure unit fixing it to one year.\\
We consider for $t$ the value provided by the last WMAP
 observations, about $13.7^{+0.2}_{-0.2} Gyr$ at 1-$\sigma$ level \cite{wmap}.\\
By this calculation we obtain an estimate of
$A^\prime_{today}\simeq7500000$. Remembering the definition of
$A^\prime$ as $A/K^{1/2}$ we can find a relation between today
scale factor and $K^{1/2}$:
\begin{equation}
A_{today}\sim 7.5\cdot10^6 K^{1/2},
\end{equation}

\noindent which can be compared with the estimate of the ratio
between scale factor and structures. If we want to refer to the
$1\sigma$ range of WMAP universe age we obtain an estimate for
$A^\prime$ to be
comprised between $\left.\right]7430000, 7580000\left.\right[$.\\
As a second case we can perform the same calculation considering
now the recombination value of cosmic time (i.e. the instant in
which matter has become transparent to radiation). So we can
provide a theoretical estimate by our model between inhomogeneity
size (which acts as a seed of structures) and scale factor at this time.\\
Obviously, we are supposing to consider the last scattering
surface as boundary of the causal universe. As a consequence, the
estimate of $A$ size corresponds to Hubble radius estimate at a
certain time. In this sense it is possible to use again the far
away limit and thus Eq.(\ref{test-time}) for
the age evaluation.\\
Now, $t_{recomb}$ is attested to the value of $379^{+8}_{-7}$kyr
as deduced by WMAP data \cite{wmap}. By the numerical integration
of Eq.(\ref{test-time}):

\begin{equation}
A^\prime_{recomb}=A_{recomb}/K^{1/2}\sim{6863}
\end{equation}

\noindent or more exhaustively, running into the 1-$\sigma$ errors
for $t_{recomb}$ we have $A^\prime\in\left.\right]6780,
6960\left.\right[$. It is evident that the inhomogeneity decreased
in time.

\section{Summary}

In this paper, we studied a model for the evolution of the
universe in presence of a spherically symmetric inhomogeneity.
This inhomogeneity has been introduced in a FRW-like metric by a
nontrivial dependence on $r$ of the scale factor and by a
non-diagonal element in the metric. A similar approach has been
already considered in \cite{peebles0} considering a diagonal
metric in absence of cosmological constant. This non-diagonal
element does introduce in the model a characteristic scale
$A_{Crit}$ ($\sim K^{1/2}$) for $A$ and a breaking of the time
reversal  $T$ and of
parity $%
P$ symmetries, which are absent in the homogeneous cosmological
models. We
consider, as source, a dust energy-momentum tensor.\\
Then we solved the exact Einstein equation separately for large
$r$ and small $r$. We showed that for $r$ and $t$ large a FRW-like
behaviour is recovered. On the other side for small $r$, thanks to
the non-diagonal term, universe dynamics is strongly affected by
inhomogeneities, in particular the scale factor does decrease with
time. The decreasing in size of the inhomogeneity could give a
natural mechanism for the structures formation without the need of
an inflationary era. Besides, this model shows an interesting
connection between the mechanism of structures formation and the
breaking of the time reversal (the, so called, ``arrow of time'')
and of the parity, in other words they have the same physical
origin. This result has been obtained without the inclusion of the
cosmological
constant or of any other kind of negative pressure.\\
To complete the study we have proposed an analysis of the model in
relation to cosmological constant. It is obtained that a
$\Lambda$-term, unlike in the standard linearized theory, does not
affect inhomogeneity dynamics preserving structures formation. On
the contrary such a term induces a significant influence on the
far away evolution driving
towards a de Sitter-like expansion.\\
A simple test on the capability of the age prediction has been
performed, obtaining an estimate of scale factor ratio versus
inhomogeneity characteristic size. This calculation has been
performed both for
today and recombination values of universe time.\\
Of course, there are many questions to be clarified. It would be
important to know how general are the characteristic features of
the model and how it depends on our technical assumptions (such as
our choice $g_{tr}\sim \frac{1}{A}$). Besides, one should have a
more rigorous analysis of the model to test the roboustness and
the stability of the results. Alternatively, it would be preferred
to have numerical solutions able to fit the proposed asymptotic
behaviours (some preliminar calculations are
quite encouraging).\\
The last interesting question we want to mention here, is the
study of the QFT on a non $P$-invariant and non $T$-invariant
background. In fact, there could be observable consequences for
the explanation of the matter-antimatter asymmetry in the study of
the creation of particles by the time evolution of the universe.

\section*{\normalsize\bf Acknowledgement}

F. Canfora and A. Troisi are very grateful to the invaluable
suggestions and the encouragements provided to them by Prof. G.
Vilasi and Dr. S. Capozziello during their work. The authors want
to thank also their friends V. F. Cardone and S. Carloni for the
useful discussions on the topics.


\begin{thebibliography}{99}


\bibitem{peebles} P.J.E. Peebles {\it Principle of Physical Cosmology},\\
Princeton Univ. Press, Princeton (1993).
\bibitem{perlmutter}  S.  Perlmutter et al.  \apj {\bf 483}, 565 (1997).
    \\ S. Perlmutter et al. {\it Nature} {\bf 391}, 51
    (1998).\\ S. Perlmutter et al. \apj {\bf 517}, 565 (1999).
\bibitem{riess} B.P. Schmidt et al. \apj {\bf 507}, 46
    (1998).\\ A.G. Riess et al. \apj {\bf 116}, 1009
    (1998).
\bibitem{cobe} G.F. Smoot, SLAC Beam Line {\bf 23N3} 2, (1993);
C. L. Bennet et al. \apj {\bf 464}, L1 (1996); A. H. Jaffe et al
\prl {\bf 86}, 3475 (2000).
\bibitem{boomerang} P. de Bernardis et al. {\it Nature} {\bf 404}, 955 (2000).
\bibitem{maxima} A. Balbi et al.\apj {\bf 558}, L145-L146 (2001);
R. Stompor et al. \apj {\bf 561},  L7-L10 (2001).

\bibitem{wmap} C.L. Bennet et al. (WMAP collaboration) astro-ph/0302207 (2003).\\ D.N. Spergel et
al.(WMAP collaboration) astro-ph/0302209 (2003).
\bibitem{holland} S. Holland, R. M. Wald, \grg {\bf 34}, 2043
(2002).
\bibitem{nonhomocosmologies} P. Coles, F. Lucchin, {\it Cosmology: The Origin and Evolution of Cosmic
Structure}, Wiley (2002).
\bibitem{krasinsky} A. Krasinsky {\it Inhomogeneous cosmological models}, Cambridge university press,
(1997).
\bibitem{peebles0} P.J.E. Peebles \apj {\bf 147}, 859 (1967).
\bibitem{t-b} R.C. Tolman {\it Relativity, Thermodynamics and
Cosmology}, Oxford Clarendon Press, (1934).
\bibitem{swiss-cheese} K. Lake, \apj {\bf 240}, 744 (1980) and references citated therein.
\bibitem{ellis} G.F.R. Ellis, H. van Elst, gr-qc/9812046.
\bibitem{swiss-cheese1} C. Bona, J. Stela, \pr {\bf D 36}, 2915
(1987) and references citated therein.
\bibitem{swiss-cheese2} H. Kozaki, K. I. Nakao, \pr {\bf D 66},
104008 (2002) and references citated therein.
\bibitem{gibbons} G. Gibbons, \pl {\bf B 84}, 431 (1982).
\bibitem{parker} L. Parker, \pr {\bf 183}, 1057 (1969).
\bibitem{klinkhamer} F. R. Klinkhamer, Nucl. Phys. {\bf B 578}, 277 (2000).
\bibitem{kolb} E.W. Kolb and  E.W. Turner {\it The Early Universe} \\
Addison-Wesley,  Redwood City, CA (1990).






\end{thebibliography}
\end{document}